\documentclass{Interspeech2024}

\usepackage{graphicx}
\usepackage{array}
\usepackage{multirow}
\usepackage{comment}
\usepackage{lipsum}  
\usepackage{url}
\usepackage{cite}

\newcommand{\joan}[1]{ {\color{red} #1} }

\newcommand{\yigit}[1]{ {\color{blue} #1} }

\newcommand{\citey}[1]{\,\cite{#1}}

\newcommand{\Table}[1]{Table~\ref{#1}}

\usepackage[x11names,dvipsnames,svgnames,table]{xcolor}
\newcommand{\cvb}[1]{{\color{RedOrange} #1}}
\newcommand{\cbb}[1]{{\color{Brown} #1}}
\newcommand{\grayrule}{\arrayrulecolor{black!20}\midrule\arrayrulecolor{black}}
\newcommand{\subsect}[1]{\vspace{0.1cm}\noindent\textbf{#1 ---}}

\interspeechcameraready

\title{A Comprehensive Real-World Assessment of Audio Watermarking Algorithms: Will They Survive Neural Codecs?}

\name[affiliation={1}{*}]{Yigitcan}{Özer}
\name[affiliation={1}{*}]{Woosung}{Choi}
\name[affiliation={1}]{Joan}{Serrà}
\breakauthors
\name[affiliation={1}]{Mayank Kumar}{Singh}
\name[affiliation={1}]{Wei-Hsiang}{Liao}
\name[affiliation={1,2}]{Yuki}{Mitsufuji}

\address{
  \textsuperscript{1}Sony AI ~~~\textsuperscript{2}Sony Group Corporation
}
\email{yiitozer@nii.ac.jp, woosung.choi@sony.com}


\begin{document}

\maketitle
\renewcommand{\thefootnote}{\fnsymbol{footnote}}
\footnotetext[1]{Equal contribution.}
\renewcommand{\thefootnote}{\arabic{footnote}}

\begin{abstract}    
%
We present the Robust Audio Watermarking Benchmark \protect\mbox{(RAW-Bench)} to foster the evaluation of deep learning-based audio watermarking algorithms, establishing a standardized benchmark and allowing systematic comparisons.
%
To simulate real-world usage, we introduce a comprehensive audio attack pipeline featuring various distortions such as compression, background noise, and reverberation and propose a diverse test dataset, including speech, environmental sounds, and music recordings.
By assessing the performance of four existing watermarking algorithms on our framework, two main insights stand out: (i) neural compression techniques pose the most significant challenge, even when algorithms are trained with such compressions; and (ii) training with audio attacks generally improves robustness, although it is insufficient in some cases. Furthermore, we find that specific distortions, such as polarity inversion, time stretching, or reverb, seriously affect certain algorithms.
Our contributions strengthen the robustness and perceptual assessment of audio watermarking algorithms across a wide range of applications while ensuring a fair and consistent evaluation approach.
The evaluation framework, including the attack pipeline, is accessible at \url{github.com/SonyResearch/raw_bench}.
\end{abstract}

\section{Introduction}
Recent advances in audio-based applications have enabled seamless content sharing, improved creative workflows, and facilitated the widespread adoption of generative AI models\citey{RenEtAl21_Fastspeech2_ICLR, LeEtAl23_Voicebox_NeurIPS, CopetEtAl23_MusicGen_NeurIPS, WuEtAl24_MusicControlNet_TASLP}.
However, such advancements have also introduced challenges in content authenticity and copyright protection\citey{WangY21_SpoofingCountermeas_INTERSPEECH, BatlleRocaEtEl24_AssessingDataRepl_ISMIR}.
To address these challenges, audio watermarking has gained attention, embedding imperceptible but detectable information into signals.
%
%
It embeds a hidden message within a carrier signal, ensuring inaudibility while enabling reliable detection and extraction\citey{HuaEtAl16_WatermarkingReview_SP}.
Recent deep learning-based methods\citey{RomanEtAl24_AudioSeal_ICML, SinghTLM24_SilentCipher_INTERSPEECH, LiuEtAl24_TimbreWatermarking_NDSS, ChenEtal23_WavMark_arXiv, OReillyEtAl24_MaskmarkNeuralWatermarking_ICASSP, SanRomanEtAl25_LatentWatermarkingAudioGen_ICASSP} have demonstrated remarkable improvements in robustness, imperceptibility, and efficiency over traditional approaches.

The effectiveness of an audio watermarking system is commonly evaluated using three criteria: \emph{imperceptibility}, \emph{robustness}, and \emph{capacity}.
Imperceptibility refers to the fidelity of the watermarked signal, which ensures that the embedded watermark remains inaudible.
Robustness refers to the successful detection of the watermark, even under distortions or attacks that degrade the carrier signal and/or the watermark.
%
Capacity represents the amount of information (that is, the number of message bits per unit time) that can be embedded in the carrier signal.
A key challenge in audio watermarking lies in the inherent trade-offs among these three properties, as optimizing one often comes at the expense of the others\citey{AgarwalSS23_SurveyRobustImpWatermarking_MTA}.

We propose the Robust Audio Watermarking Benchmark \protect\mbox{(RAW-Bench)}, focusing on imperceptibility and robustness, with comparable capacity across models.
%
In this benchmark, we assume a threat model where adversaries have access only to the audio file, while the watermarking methods remain hidden. However, they can manipulate the audio file to prevent detection of the embedded watermark. For example, one might compress and then decompress the audio to damage the imperceptible watermark while maintaining the perceptible quality.

Our analysis evaluates four publicly available pre-trained baseline models (AudioSeal\citey{RomanEtAl24_AudioSeal_ICML}, SilentCipher\citey{SinghTLM24_SilentCipher_INTERSPEECH}, Timbre\citey{LiuEtAl24_TimbreWatermarking_NDSS}, and WavMark\citey{ChenEtal23_WavMark_arXiv}) under various distortions, including mixing, background noise, filtering, reverberation, compression, and equalization. 
We also study the impact on the robustness of distortion-aware training, integrating a comprehensive audio attack pipeline into the training of AudioSeal and SilentCipher.
To systematically evaluate performance, we introduce a novel test dataset covering multiple audio domains, including music, speech, and environmental sounds, with non-compressed raw recordings.
Our findings provide two main insights: (i) neural compression (e.g.,\,Encodec\citey{DefossezCSA23_Encodec_TMLR} and Descript Audio Codec\citey{KumarEtAl23_DAC_NeurIPS}) poses the greatest challenge to audio watermarking systems, even when these systems are trained with such compressions; and (ii) training with audio attacks improves robustness, consistent with observations by Juvela and Wang\citey{JuvelaW25_CodecAugmentationWatermarking_ICASSP}, although it does not guarantee good performance.
Additionally, we observe that specific distortions, such as polarity inversion, time stretching, and reverb, severely impact certain watermarking algorithms. We end with a discussion of future perspectives regarding the trade-off between audio watermarking and neural compression.

\begin{table*}[t]
    \centering
    \caption{Characteristics of baseline watermarking models. Capacity corresponds to the one we set/use for our evaluation.}
    \resizebox{1\textwidth}{!}{
    \begin{tabular}{llccccl}
        \toprule
        \textbf{Model} & \textbf{Domain} & \textbf{Sample rate (kHz)} & 
        \textbf{Message (bits)} & \textbf{Capacity (bps)} & \textbf{Training size (h)} & \textbf{Training data} \\
        \midrule
        AS: AudioSeal\citey{RomanEtAl24_AudioSeal_ICML} & Waveform & 16 & 16 & 5.33 & 4500 & Speech \\
        SC: SilentCipher\citey{SinghTLM24_SilentCipher_INTERSPEECH} & Spectral & 16 & 23.8 & 5.33 & 372 & Speech, music, TV shows \\
        TI: Timbre\citey{LiuEtAl24_TimbreWatermarking_NDSS} & Spectral & 22.05 & 30 & 5.00 &  100 & Speech \\
        WM: WavMark\citey{ChenEtal23_WavMark_arXiv} & Spectral & 16 & 16 & 5.28 & 5000 & Speech, music, environmental \\
        \bottomrule
    \end{tabular}
    }
    \label{tab:baseline_models}
\end{table*}

\begin{table*}[t]
\centering
\caption{Audio attack pipeline for robustness analysis. The threshold is used to separate between loose (L) and strict (S) attacks.}
\resizebox{0.87\textwidth}{!}{%
\begin{tabular}{lllcc}
\toprule
\textbf{Attack Category}        & \textbf{Attack Type}              & \textbf{Parameter}      & \textbf{Range} &  \textbf{L/S Threshold} \\ \midrule
\multirow{3}{*}{Mixing}       & GN: Gaussian noise              & SNR (dB)                                   & {[}20, 60{]}         & 40                              \\ 
                              & BN: Background noise (from\citey{ThiemannIV13_DEMAND_PMA})           & SNR (dB)                                   & {[}20, 60{]}         & 35                              \\ 
                              & RV: Reverb (from\citey{JeubSV09_AachenImpulseResponseDatabase_DSP})                     & SNR (dB)                                   & {[}0, 12{]}          & 6                               \\ \midrule
\multirow{3}{*}{Dynamics}     & DC: Dynamic range compression  & Threshold (dB)                              & {[}$-$36, $-$6{]}    & $-$18                               \\ 
                              & DE: Dynamic range expansion    & Threshold (dB)                              & {[}$-$16, $-$6{]}    & $-$12                               \\ 
                              & LM: Limiter                     & Threshold (dB)                             & {[}$-$36, $-$6{]}    & $-$18                               \\ \midrule
\multirow{3}{*}{Filtering}    & LP: Lowpass                     & Cutoff (Hz)                                & {[}3500, 8000{]}     & 6000                            \\ 
                              & HP: Highpass                    & Cutoff (Hz)                                & {[}10, 500{]}        & 250                             \\ 
                              & EQ: Equalization                & Max gain (dB)                              & {[}$-$0.75, 0.75{]}  & $\pm$ 0.375     \\ \midrule
\multirow{6}{*}{Low level}    & TS: Time stretch                & Rate                                       & {[}0.75, 1.25{]}     & 1.00 $\pm$ 0.05        \\ 
                              & TJ: Time jittering              & Scale                                      & {[}0.10, 0.50{]}       & 0.20                             \\ 
                              & PI: Polarity inversion          & N/A                                        & N/A                  & N/A                             \\ 
                              & GA: Gain adjustment             & Rate                                       & {[}0.20, 5{]}         & 1.00 $\pm$ 0.50                  \\ 
                              & QN: Quantization                & \#Bits/sample                                     & \{8, 9, ... 16\}     & 12                                \\ 
                              & PS: Phase shift                 & Seconds                                & {[}-0.10, 0.10{]}                 & 0 $\pm$ 0.05                             \\ \midrule
\multirow{2}{*}{Neural compression} & EN: Encodec\citey{DefossezCSA23_Encodec_TMLR} (at 24\,kHz)      & \#Codebooks       & \{16, 32\}           & 32        \\ 
                              & DA: Descript Audio Codec\citey{KumarEtAl23_DAC_NeurIPS} (at 44.1\,kHz)  & \#Codebooks       & \{7, 8, 9\}          & 9         \\ \midrule
\multirow{3}{*}{Conventional compression} & MP: MP3 codec             & Bitrate (kbps)                             & \{64, 128, 256\}     & 64                              \\ 
                              & OG: OGG codec                        & Bitrate (kbps)                             & \{48, 64, 128, 256\} & 48                              \\ 
                              & AA: AAC codec                        & Bitrate (kbps)                             & \{64, 128, 256\}     & 64                              \\ \bottomrule
\end{tabular}
}
\label{tab:audio_attacks}
\end{table*}

\section{Related Work}
\label{sec:related_work}
To the best of our knowledge, the only study that compares deep learning-based audio watermarking models is AudioMarkBench\citey{LiuEtAl24_AudioMarkBench_NeurIPS}. 
AudioMarkBench is a benchmarking framework that evaluates the robustness of three audio watermarking models (AudioSeal, Timbre, and WavMark), using their publicly available pre-trained weights, on a subset of speech signals sampled at 16\,kHz.
Apart from additionally considering the imperceptibility criterion, our work diverges from AudioMarkBench in a number of important aspects.
First, instead of relying on compressed recordings, we construct a high-fidelity test dataset containing raw, non-compressed audio at 44.1\,kHz. 
This avoids any potential confounding factor introduced by low-bandwidth or compressed signals.
Second, instead of focusing on speech, we base our results on a selection of speech, music, and environmental sounds. 
This ensures a more comprehensive evaluation under a variety of real-world signals.
Third, we extend the analysis by including SilentCipher, another baseline approach with competitive performance. This broadens the scope of our work. 
Fourth, we study the effect of retraining watermarking algorithms using the proposed audio attack pipeline. 
This allows us to isolate the impact of training-time distortions on watermark robustness, providing deeper insights into the advantages of training with adversarial attacks. 
Finally, it is also worth mentioning that our attack/test pipeline is larger and more varied than the one of AudioMarkBench (20 vs.\ 12 distortions, respectively, see below).
%

%
All models considered in this paper encode a hidden message in a mono-carrier signal, but differ in architecture, sampling rate, training dataset, and operating domain, as detailed in~\Table{tab:baseline_models}. 
To ensure an unbiased evaluation, we verify that no training data from the considered models is included in our test set. 
This guarantees that our analysis and conclusions are based entirely on out-of-sample data. 
AudioSeal (AS)\citey{RomanEtAl24_AudioSeal_ICML} is originally trained with various distortion augmentations, including time modifications, filtering, audio effects, and compression.
%
%
SilentCipher (SC)\citey{SinghTLM24_SilentCipher_INTERSPEECH} considers time jittering, additional noise, and non-differentiable compression techniques as augmentations.
Additionally, it introduces a lower SDR bound on the watermarked signals to account for imperceptibility.
Timbre (TI)\citey{LiuEtAl24_TimbreWatermarking_NDSS} incorporates ISTFT, normalization, transformation, and wave reconstruction for robustness.
WavMark (WM)\citey{ChenEtal23_WavMark_arXiv} employs a curriculum learning strategy and applies various distortions during training, including noise, filtering, compression, echo, and time stretching.

\section{Methodology}
\label{sec:methodology}


\subsect{Test Dataset} To evaluate watermarking algorithms in various domains, we create a comprehensive test dataset using open-source collections from various sources.
It includes classical and popular music, speech, and environmental sounds, which account for a wide range of real-world use cases.
To maintain a high fidelity, all audio recordings in the dataset have sample rates equal to or exceeding 44.1\,kHz, and they are provided as raw, non-compressed files.
Our test dataset is formed by the union of the following publicly-available collections\footnote{\url{github.com/SonyResearch/raw_bench}}:

\begin{itemize}
    \item Bach10\citey{DuanP11_SoundPrism_JSTSP} -- A dataset of ten classical ensemble recordings.
    \item Clotho\citey{DrossosLV20_Clotho_ICASSP} -- A collection of diverse environmental sounds.
    \item Device and Produced Speech (DAPS)\citey{Mysore15_DAPS_SPS} -- A dataset of studio-quality speech recordings alongside consumer-device recordings captured in real-world environments.
    \item FreiSchuetz\citey{PraetzlichMBV15_FreiDi_ISMIR-LBD} -- A dataset of professional stereo mixes and raw multitrack recordings of three opera performances.
    \item GuitarSet\citey{XiEtAl18_GuitarSet_ISMIR} -- A dataset of solo guitar recordings.
    \item jaCapella\citey{NakamuraEtAl23_jaCapella_ICASSP} -- A corpus of 50 Japanese a cappella vocal ensemble recordings, including individual voice parts.
    \item MAESTRO\citey{HawthorneSRSHDE19_MAESTRO_ICLR} -- A dataset of paired audio and MIDI recordings from the International Piano-e-Competition.
    \item MoisesDB\citey{PereiraAKV23_MoisesDB_ISMIR} -- A dataset of 240 musical tracks spanning twelve genres, performed by 45 artists (we use only the mixes).
    \item Piano Concerto Dataset (PCD)\citey{OezerSALSM23_PCD_TISMIR} -- 
    A collection of piano concerto excerpts (we use only the raw piano tracks).
\end{itemize}

\subsect{Attack Pipeline} For the robustness analysis, we develop a comprehensive audio attack pipeline that simulates 20 real-world distortions (\Table{tab:audio_attacks}). 
The distortions are organized into six categories and are designed to simulate real-world variability in playback and processing. 
As many attacks allow for parameter variation (e.g.,\,different levels of noise, filtering, or compression strength), we first establish a range of suitable values that align with real-world conditions (\Table{tab:audio_attacks}, Range).
Next, we consider two settings based on the strength of such parameters: \emph{loose} and \emph{strict}.
%
%
The loose setting corresponds to imperceptible distortions, while the strict setting represents cases where distortions are audible but still acceptable. 
To define a threshold between these two settings (\Table{tab:audio_attacks}, Threshold), we conducted an internal listening test with five expert listeners who evaluated the perceptibility and acceptability of each attack. 
The threshold was then set based on whether they could perceive a notable subjective difference compared to the original audio.
For example, the listeners unanimously agreed that Gaussian noise with Signal-to-Noise Ratio (SNR) values in the range of {[}40, 60{]} dB is almost imperceptible, placing it in the loose class.
In contrast,  noise in the range of {[}20, 40{]} dB is audible but still acceptable, categorizing it under the strict class.
These thresholds allow us to systematically evaluate watermarking models under both realistic (loose) and challenging (strict) conditions, ensuring a meaningful robustness analysis.

Building on this attack framework, our robustness experiments follow a two-stage process.
First, we evaluate pre-trained models by assessing their performance against the full set of distortions (loose and strict setups).
Second, we retrain AS and SC using our audio attack pipeline under the strict parameter settings to examine the impact of adversarial training on watermark robustness.
To ensure a balanced exposure to different distortions, we employ a uniform weighting scheme per attack category, along with spectrogram augmentation\citey{ParkEtAl19_SpecAugment_INTERSPEECH}.
For retraining, we utilize a proprietary dataset consisting of approximately 1250 hours of musical mixes, along with 40 hours of VCTK\citey{VeauxYM19_VCTK} and a 40-hour subset of environmental sounds from BBC Sound Effects\citey{BBCSoundEffects91}.

\subsect{Evaluation Metrics} To evaluate the performance of the considered methods, we employ a set of metrics that measure either the robustness of watermark detection or the imperceptibility of the watermark. 
Our analysis focuses on these two aspects while keeping the capacity constant and comparable across all considered models (\Table{tab:baseline_models}).
For imperceptibility, we consider:
\begin{itemize}
\item Scale-invariant signal-to-noise ratio (SI-SNR)\citey{LuoM19_ConvTasNet_TASLP} -- SI-SNR measures the distortion or noise in a processed signal relative to a reference, independent of scaling.
\item Mel cepstral distance (MCD)\citey{Kubichek93_MelCepstralDist_PACRIM} -- MCD is a perceptually motivated measure that quantifies the spectral difference between the original and the watermarked audio.
%
%
%
\item Virtual Speech Quality Objective Listener (MOS-LQO)\citey{HinesSKH15_ViSQOL_EUROSIP} -- An objective, full-reference metric for assessing perceived audio quality based on spectro--temporal similarity.
%
%
\end{itemize}
For robustness, we use:
\begin{itemize}
\item Bitwise accuracy -- This metric measures the proportion of correctly decoded bits in the detected watermark message. 
\item Message accuracy -- This metric assesses the overall success of the watermark extraction process (that is, if all bits in the extracted message match the original watermark).
\end{itemize}

\section{Results and Discussion}
\label{sec:exp}
%
\subsect{Imperceptibility} As a first step, we evaluate the considered models in clean (distortion-free) conditions, focusing on overall perceptual quality and detection accuracy (\Table{tab:perceptual_evaluation}).
%
%
%
Among all pre-trained models, SC consistently outperforms others in perceptual quality, achieving the highest SI-SNR and lowest MCD, indicating minimal perceptual impact from watermark insertion.
Similarly, SC achieves the highest MOS-LQO score, closely followed by AS, while TI performs the worst across all perceptual metrics.
The better performance of SC regarding the perceptual metrics can be attributed to its lower SDR bound constraint of the watermarked signals.
In terms of overall robustness, all models achieve accuracies close to 1 in clean conditions (\Table{tab:perceptual_evaluation}, ACC), and we additionally measure a true-positive rate between 0.97 and 1 at zero false-negative rate for all of them (not shown). Overall, this confirms a reliable watermark extraction in the absence of audio attacks.
In this clean setup, results for the re-trained models AS$^{*}$ and SC$^{*}$ do not differ much from the pre-trained ones, except for the case of SC$^{*}$ with SI-SNR and MCD, which we on purpose re-train with a lower SDR bound to improve robustness (see below).

\begin{table}[t]
\centering
\caption{Comparison of models across different metrics on clean watermarked audio (no attacks). ACC shows average bit-wise/full-message accuracy, and an asterisk ($^{*}$) indicates that the model has been retrained with the strict attacks.}
\resizebox{\columnwidth}{!}{%
    \begin{tabular}{lcccc}
    \toprule
    \textbf{Model} & \textbf{SI-SNR} $\uparrow$ & \textbf{MCD} $\downarrow$ & \textbf{MOS-LQO} $\uparrow$ & \textbf{ACC} $\uparrow$ \\
    \midrule
    AS & 22.73 & 0.53 & 4.93 & 0.997 / 0.962 \\ 
    SC & 49.13 & 0.25 & 4.98 & 0.999 / 0.993 \\
    TI & 21.91 & 1.74 & 4.59 & 1.000 / 1.000 \\
    WM & 35.89 & 0.62 & 4.91 & 0.998 / 0.993 \\
    \midrule
    AS$^{*}$ & 23.60 & 0.49 & 4.95 & 0.999 / 0.997 \\    
    SC$^{*}$ & 31.80 & 1.04 & 4.88 & 0.999 / 0.993 \\
    \bottomrule
    \end{tabular}%
}
\label{tab:perceptual_evaluation}
\end{table}

\begin{table}[t]
\centering
\caption{Average bitwise/full-message accuracy across all strict attacks, for different audio domains.}
\resizebox{0.7\columnwidth}{!}{%
\begin{tabular}{lccc}
\toprule
\textbf{Model} & \textbf{Environ.} & \textbf{Music} & \textbf{Speech} \\
\midrule
AS & .91 / .68 & .91 / .68 & .91 / .72 \\
SC & .73 / .47 & .75 / .49 & .81 / .62  \\
TI &  .94 / .79 & .93 / .78 & .94 / .78 \\
WM & .74 / .70 & .77 / .72 & .80 / .77 \\
\midrule
AS$^{*}$ & .95 / .81 & .95 / .81 & .94 / .80  \\
SC$^{*}$ & .91 / .75 & .90 / .79 & .92 / .80 \\
\bottomrule
\end{tabular}
}
\label{tab:comparison_domain}
\vspace{-0.2cm}
\end{table}

\subsect{Audio Domains} Next, we analyze the robustness across the different audio domains found in our test dataset (\Table{tab:comparison_domain}).
We find that all models exhibit similar performance across environmental sounds, music, and speech, with only minor variations in accuracy between domains.
Interestingly, AS and TI were trained exclusively on speech data, yet they generalize well to the other two domains.
%
%
%
This suggests that the considered models can generalize well across different audio domains.
We also observe that training with adversarial attacks further improves robustness, especially for full-message accuracy.

\begin{table*}[th]
\centering
\caption{Comparison of bitwise (top) and full-message (bottom) robustness for the considered models under various attacks (columns, see abbreviations in Table~\ref{tab:audio_attacks}). For each model, we evaluate the strict (S) and loose (L) settings (Eval column). An asterisk ($^{*}$) indicates that the model has been retrained with the strict attacks, and a checkmark (\checkmark) indicates an accuracy of 0.99 or above.}
\resizebox{\textwidth}{!}{%
\begin{tabular}{>{\raggedleft\arraybackslash}ll
ccccccccccccccccccccc}
\toprule
\textbf{Model} & \textbf{Eval} & \textbf{GN} & \textbf{BN} & \textbf{RV} & \textbf{DC} & \textbf{DE} & \textbf{LM} & \textbf{LP} & \textbf{HP} & \textbf{EQ} & \textbf{TS} & \textbf{TJ} & \textbf{PI} & \textbf{GA} & \textbf{QN} & \textbf{PS} & \textbf{EN} & \textbf{DA} & \textbf{MP} & \textbf{OG} & \textbf{AA} \\
\midrule
AS&S&\checkmark&\checkmark&.87&\checkmark&\checkmark&.98&\checkmark&.96&.91&.97&\checkmark&\cvb{.18}&\checkmark&\checkmark&\cbb{.62}&.96&\cbb{.52}&\checkmark&\checkmark&\checkmark \\
SC&S&\cbb{.63}&.98&.80&.96&.91&.83&\checkmark&.93&.92&\cvb{.41}&.86&\checkmark&\checkmark&\checkmark&\cbb{.63}&\cvb{.33}&\cvb{.32}&\cbb{.54}&\cbb{.61}&.93 \\
TI&S&.98&\checkmark&.96&\checkmark&\checkmark&\checkmark&\checkmark&\checkmark&\checkmark&\checkmark&\checkmark&\checkmark&\checkmark&\checkmark&\checkmark&\cbb{.64}&\cbb{.60}&\checkmark&\checkmark&\checkmark \\
WM&S&.83&\checkmark&.89&.98&.95&.95&\checkmark&\checkmark&\checkmark&.82&.98&\checkmark&.98&\checkmark&\checkmark&\cvb{.00}&\cvb{.00}&\cvb{.42}&.89&\checkmark
\\
\grayrule
AS&L&\checkmark&\checkmark&.97&\checkmark&\checkmark&\checkmark&\checkmark&\checkmark&\checkmark&.97&\checkmark&\cvb{.18}&\checkmark&\checkmark&\cbb{.60}&.97&\cbb{.53}&\checkmark&\checkmark&\checkmark \\
SC&L&\checkmark&\checkmark&.89&\checkmark&.97&\checkmark&\checkmark&\checkmark&\checkmark&\cbb{.64}&.88&\checkmark&\checkmark&\checkmark&\cbb{.65}&\cvb{.33}&\cvb{.32}&\cbb{.57}&\cbb{.65}&.98
 \\
TI&L&\checkmark&\checkmark&.98&\checkmark&\checkmark&\checkmark&\checkmark&\checkmark&\checkmark&\checkmark&\checkmark&\checkmark&\checkmark&\checkmark&\checkmark&\cbb{.65}&\cbb{.62}&\checkmark&\checkmark&\checkmark
\\
WM&L&\checkmark&\checkmark&.98&\checkmark&\checkmark&\checkmark&\checkmark&\checkmark&\checkmark&.95&\checkmark&\checkmark&.98&\checkmark&\checkmark&\cvb{.00}&\cvb{.00}&\cvb{.48}&.83&\checkmark
\\
\grayrule
AS$^{*}$&S&\checkmark&\checkmark&.91&\checkmark&\checkmark&\checkmark&\checkmark&\checkmark&.96&\checkmark&\checkmark&.98&\checkmark&\checkmark&\cbb{.62}&.97&\cbb{.60}&\checkmark&\checkmark&\checkmark 
\\
SC$^{*}$&S&\checkmark&\checkmark&\checkmark&\checkmark&\checkmark&\checkmark&\checkmark&\checkmark&\checkmark&\cbb{.54}&\checkmark&\checkmark&\checkmark&\checkmark&.91&\cbb{.67}&\cvb{.42}&.98&\checkmark&\checkmark \\
AS$^{*}$&L&\checkmark&\checkmark&\checkmark&\checkmark&\checkmark&\checkmark&\checkmark&\checkmark&\checkmark&\checkmark&\checkmark&.98&\checkmark&\checkmark&\cbb{.59}&\checkmark&\cbb{.61}&\checkmark&\checkmark&\checkmark
\\
SC$^{*}$&L&\checkmark&\checkmark&\checkmark&\checkmark&\checkmark&\checkmark&\checkmark&\checkmark&\checkmark&.84&\checkmark&\checkmark&\checkmark&\checkmark&.92&.75&\cvb{.44}&.98&\checkmark&\checkmark \\
\midrule
AS&S&.93&.95&\cvb{.22}&.92&.89&.88&.93&\cbb{.57}&\cvb{.39}&\cbb{.72}&.96&\cvb{.00}&.91&.95&\cvb{.06}&\cbb{.65}&\cvb{.00}&.92&.95&.96 \\
SC&S&\cvb{.29}&.93&\cvb{.45}&.78&\cbb{.64}&\cvb{.49}&\checkmark&\cbb{.71}&\cbb{.69}&\cvb{.00}&\cbb{.60}&.98&\checkmark&\checkmark&\cvb{.24}&\cvb{.00}&\cvb{.00}&\cvb{.14}&\cvb{.22}&.75 \\
TI&S&.80&.98&\cbb{.57}&.96&.94&.95&\checkmark&\checkmark&\checkmark&.90&.97&\checkmark&\checkmark&\checkmark&\checkmark&\cvb{.00}&\cvb{.00}&.92&.97&\checkmark
\\
WM&S&\cbb{.74}&.98&\cbb{.73}&.95&.90&.90&\checkmark&\checkmark&.98&\cbb{.60}&.96&\checkmark&.94&\checkmark&\checkmark&\cvb{.00}&\cvb{.00}&\cvb{.29}&.79&\checkmark \\
\grayrule
AS&L&.95&.96&\cbb{.72}&.96&.94&.96&.96&.88&.90&\cbb{.71}&.95&\cvb{.00}&.91&.95&\cvb{.02}&.78&\cvb{.00}&.93&.95&.96 \\
SC&L&\checkmark&\checkmark&\cbb{.67}&\checkmark&.89&.97&\checkmark&.96&.96&\cvb{.20}&\cbb{.61}&.98&\checkmark&\checkmark&\cvb{.28}&\cvb{.00}&\cvb{.00}&\cvb{.17}&\cvb{.30}&.94 \\
TI&L&\checkmark&\checkmark&.77&\checkmark&.98&\checkmark&\checkmark&\checkmark&\checkmark&.92&.97&\checkmark&.98&\checkmark&\checkmark&\cvb{.00}&\cvb{.00}&.94&.94&\checkmark
\\
WM&L&\checkmark&\checkmark&.93&\checkmark&.98&\checkmark&\checkmark&\checkmark&\checkmark&.88&.98&\checkmark&.97&\checkmark&\checkmark&\cvb{.00}&\cvb{.00}&\cvb{.34}&.75&\checkmark
\\
\grayrule
AS$^{*}$&S&.98&\checkmark&\cvb{.41}&\checkmark&\checkmark&\checkmark&\checkmark&.98&\cbb{.62}&.95&\checkmark&\cbb{.72}&\checkmark&\checkmark&\cvb{.03}&.79&\cvb{.00}&\checkmark&\checkmark&\checkmark
\\
SC$^{*}$&S&.96&.98&.95&\checkmark&.98&.98&.98&.97&.97&\cvb{.00}&\checkmark&.98&\checkmark&\checkmark&\cbb{.74}&\cvb{.06}&\cvb{.00}&.85&.91&.98\\
AS$^{*}$&L&\checkmark&\checkmark&.90&\checkmark&\checkmark&\checkmark&\checkmark&\checkmark&\checkmark&.93&\checkmark&\cbb{.72}&\checkmark&\checkmark&\cvb{.01}&.90&\cvb{.00}&\checkmark&\checkmark&\checkmark
\\
SC$^{*}$&L&\checkmark&\checkmark&.98&\checkmark&\checkmark&\checkmark&.98&\checkmark&.98&\cvb{.37}&\checkmark&.98&\checkmark&\checkmark&.75&\cvb{.12}&\cvb{.00}&.87&.91&.98\\
\bottomrule
\end{tabular}%
}
\label{tab:robustness}
\vspace{-0.12cm}
\end{table*}

\subsect{Robustness} We now evaluate how different distortions impact the robustness of watermarking models.
\Table{tab:robustness} presents the bitwise (top) and full-message (bottom) accuracy of both pre-trained and retrained models under strict and loose attack conditions.
In general, TI demonstrates the highest robustness, which is expected as compensation for the low imperceptibility scores obtained above. In addition, different models show varying degrees of resilience to specific attacks.
For example, AS struggles with PI and PS, SC is particularly vulnerable to TS, PS, MP, and OG, and WM is notably affected by MP compression.
Among all distortions, neural compression methods (EN and DA) pose the greatest challenge, particularly for SC and WM under strict conditions, with bitwise accuracies dropping to 0.33 and 0, respectively. 
DA has a more severe effect on all models. Full-message accuracy is unacceptable for all models and settings, except for AS with EN compression, which can be explained by the fact that the architecture of AS is based on EN.
%
%
%
However, importantly, the advantage of using the EN architecture does not extend to DA, suggesting that employing a neural codec architecture does not generalize to alternative neural codecs.
Overall, this indicates that audio watermarking models fail under neural compression, highlighting a critical weakness.

\subsect{Re-training} Now we assess the effect of incorporating adversarial attacks in SC and AS training (\Table{tab:robustness}).
While improving robustness against certain attacks (e.g., GN for SC and PI for AS), it does not fully resolve vulnerabilities in other cases: even after retraining, models continue to struggle with compression (EN, DA, MP, OG), RV, and PS.
This suggests that some attacks introduce fundamental challenges that adversarial-attack training augmentations fail to overcome.
Full-message accuracy is poor across all models, even after retraining, highlighting a fundamental limitation of existing methods.


\subsect{Will Watermarks Survive Neural Codecs?} One of the main performance gaps we observe in our results is the robustness to neural codecs. 
Bitwise accuracies are generally below 0.5, and full-message accuracies are around 0 for almost all approaches in both EN and DA. While AS performs better than the rest for EN and training with neural-codec attacks can help AS and SC, none of the considered watermarking algorithms survives the DA attack. 
Also, retraining does not bring robustness to neural codecs up to acceptable levels. 
This suggests that there is a fundamental issue underlying such poor generalization and lack of performance. 
In fact, watermarking algorithms and neural codecs compete for the same space. 
On the one hand, watermarking algorithms strive to insert imperceptible information into the audio signal, but on the other hand, neural codecs strive to remove the imperceptible information from the (possibly the same) audio signal. Deep learning methodologies have recently enhanced the capabilities of both types of algorithms. However, if we consider the limit situation where both algorithms successfully achieve their purpose, we believe that neural codecs will end up removing imperceptible watermarks. In addition, neural codecs are usually the final stage in the audio processing pipeline and thus have more chance/incentive to remove any imperceptible information, regardless of its origin.

\section{Conclusion}
\label{sec:conclusion}

We introduced a systematic evaluation framework for deep learning-based audio watermarking algorithms, addressing important gaps in robustness analysis and benchmarking. 
We designed a comprehensive audio attack pipeline that simulates real-world distortions, and introduced a diverse test dataset comprising multiple audio domains. By studying the performance of four existing watermarking algorithms within our framework, we were able to provide novel insights regarding imperceptibility and robustness to specific attacks.
On the whole, our framework contributes to the development of more resilient and perceptually optimized audio watermarking systems. We believe future work should focus on the trade-off between audio watermarking and neural codecs, which our study discusses and shows to be a critical point.

\bibliographystyle{IEEEtran}
\bibliography{bibliography}

\end{document}